# Electrical properties of vanadium oxide subject to hydrogen plasma treatment

Alexander Pergament, Nikolai Kuldin

Physics and Technology Department, Petrozavodsk State University,

Petrozavodsk, 185910, Russian Federation

e-mail: aperg@psu.karelia.ru

**Abstract.** The effect of doping with hydrogen on the electrical properties of vanadium oxide is studied. For vanadium oxide films, subject to cold hydrogen plasma treatment, the temperature dependence of resistance with a maximum at $T \sim 100$ K is observed.  Also, the dependence of the a.c. resistance on frequency is studied. A strategy for fabrication new superconducting materials is discussed.

The recent discovery of high-temperature superconductivity (HTSC) in a class of copperless materials Ln-O(F)-Fe-As (Ln = La, Ce, Sm, etc) with a maximum $T_c \geq 50$ K [1] has greatly stimulated the interest to the search for new SC /materials. Not only new HTCS materials with higher $T_c$'s (and even RTSC – "room-temperature SC", according to the Academician Ginzburg terminology [2]) underlie this interest; but also it is hoped that a discovery of any new SC materials, even those with relatively low $T_c$, would allow further understanding the problem of HTSC mechanism on the whole.

Many works [3-7] have been devoted to advocating the idea of non-Cu-based oxide HTSC. The materials suggested are, mostly, strongly correlated transition metal oxides which are close to a metal-insulator transition (MIT). This strategy is based, of course, on the known properties of the "usual" high-$T_c$ compounds (YBaCuO, etc).

It is interesting to note that iron is presented in the new layered HTSC materials (above-mentioned oxypniktides [1,8,9]), though it is known that ferromagnetism should destroy usual BCS superconductivity [8]. Moreover, HTSC oxycuprates and oxypniktides have, generally speaking, different parental ground states. The former, in the underdoped state, are, as a rule, Mott insulators, whereas the latter are (according to different data) either non-magnetic metals or antiferromagnetic semimetals [8,9]. Nonetheless, as has been argued in the work [10], "approaching the Mott state, we will find SC". This means that the SC state should inevitably be the final point of the Mott localization in the so-called "bad" metals [11].

Possible host structures might be found among those materials undergoing metal–insulator transitions as a function of temperature [3]. Materials with metallic ground states can then be generated by substitution. It has been suggested that simple binary compounds, such as titanium or vanadium oxides belonging to the homologous series $M_nO_{2n-1}$ may be good candidates for such precursor host structures [3,6,7].

Thus, as we see, HTSC-materials based on vanadium oxides (with the suppressed MIT) are actively discussed in the scientific literature [3-7]. They are studied in many labs all over the world, and even have been patented [12], though, unfortunately, not discovered yet. A "first swallow" in this intriguing area of research is presumably the experiment described in our previous work [7] in this journal. The result is presented in Fig.1. A sample of $H_xV_2O_5 \times yH_2O$ ($V_2O_5$-gel) was subject to the hydrogen plasma treatment in an RF reactor with remote low-temperature plasma. One can see that the curve has a distinctive maximum at $T = T_m \approx 100$ K. Such behaviour of the $R(T)$ dependence might be attributed to an inverse (or re-entrant) MIT [7], which is characteristic of, for instance, CMR-manganites [13] (cf. curves 1 and 2 in Fif.1, a).

However, this might also be an indication of the SC behaviour: similar dependence of $R$ on $T$ is characteristic of many under- and over-doped HTSC cuprates (the example of such behaviour is shown in Fig.1, b). Little is known about composition and structure of the vanadium-oxide phase, obtained by the treatment in low-temperature plasma, which is responsible for the observed properties [1]. Presumably, it is some hydrogen bronze of a Magneli-phase lower oxide $V_nO_{2n-1}$. For example, it could be $H_xV_7O_{13}$ [2], which accounts for (in view of aforesaid) the possible SC properties of this specimen (with $T_c \sim 100$ K – see Fig.1, a, curve 1).

Actually, X-ray structural and phase analysis reveals a mixture of lower vanadium oxides. Multi-phase heterogeneous systems exhibit usually well-defined behaviour in frequency electrical measurements: $R \sim (f)^{0.7}$ (which also is inherent to hopping conductivity of disordered materials [16]). Actually, similar dependence has been observed for the $V_2O_5$-gel [17] which serves as a precursor for the samples described here.

The dependences of the plasma-treated sample resistance on frequency $f$ at different temperatures are presented in Fig.2. However, in the process of frequency measurements, we encountered the effect of a shift of the $T_m$ toward higher values as the frequency increases. Such a behaviour is often characteristic of relaxor ferroelectrics with frequency-induced shift of $\varepsilon(T)$ curves, as well as of conductivity in spin glasses and materials with specific magnetic phase transitions (like, e.g., quasi-crystal $AlPdMn_x$ or $CeAl_3$) [16,18,19].

Note that, as was said above, a hydrogen bronze $H_xV_nO_{2n-1}$ of a Magneli-phase lower vanadium oxide (predicted in [7] as a possible vanadium-based SC phase) might turn out to be a good candidate for HTSC. On the one hand, hydrogen leads to metallization of the initially semiconducting material (see [7] and references therein), and on the other hand, it acts as a ligand promoting SC properties. The point is that initial structures (or precursors) to produce new SC materials could be a material exhibiting a temperature-induced MIT with the metallic state, which is somehow stabilized down to T=0 K. Presumably, these should be multi-component compounds possessing a high capacity for introducing impurity (desirably, up to ~10% without the separate phase precipitation) [20]. Other components should be: i) a transition or rare-earth element (e.g., V itself or some additional element) ensuring band narrowness and a high DOS on the Fermi level, and ii) some light (H, Li, Na, etc) element. The latter demand has been clarified earlier in [7] with the reference to the same work [20]. Note that exactly the very approach is used in the search for SC in transition metal hydrides [21].

------------------

[1] The difficulties of the samples characterization have been described in [7]. Manifestly, further efforts are necessary to unravel this problem.

[2] $V_7O_{13}$ is the only of vanadium oxides remaining metallic down to liquid He temperatures [15].

Finally, we would like, instead of a formal conclusion, just to represent a brief synopsis of compounds (or systems of compounds), illustrating the above speculations concerning the intimate link between a temperature-induced MIT and SC in different transition metal oxides (see Table 1 below).

Table 1.

Metal-insulator transitions and superconductivity in several transition-metal and rare-earth oxides. (Columns 2 and 3 – MIT at $T = T_t$; columns 4 and 5 – SC transition at $T = T_c$)

| System | MIT | $T_t$ (K) | SC | $T_c$ (K) |
|---|---|---|---|---|
| Tungsten oxide bronzes | $WO_{3-x}$ [16] | 250 | $A_xWO_3$ (A=Na, K, Rb, Cs, Ca, Sr, Ba, Tl, In) [4,20] | 0.5 – 7 ~90 (for A=Na and x=0.05) [3] |
| Oxide pyrochlores | $Cd_2Os_2O_7$ [22] | 225 | $Cd_2Re_2O_7$ $KOs_2O_6$ [22] | 1 9.9 |
| Ruthenates | $Ca_2RuO_4$ $Ca_3Ru_2O_7$ [23] | 357 48 | $Sr_2RuO_4$ $RuSr_2GdCu_2O_8$ [23] | 1.5 40 |
| Ti – O | $Ti_nO_{2n-1}$, e.g. $Ti_4O_7$ [15,18] | 150 | $Li_{1+x}Ti_{2-x}O_4$ [4] | 13 |
| V – O | $V_nO_{2n-1}$, e.g. $VO_2$ [15,18] | 340 | $Tl_{0.2}VSrO_y$ [24] | ~80 (from diamagnetic signal) [3] |
| Nb – O | $NbO_2$ [14,18] | 1070 | $(Sr_{1-x}La_x)Nb_2O_{6-y}$ with x≤0.3 [25] | 11.5 |
| Co – O | $LaCoO_3$ [14,15] | ~500 | $Na_xCoO_2 \cdot yH_2O$ [26] | ~ 4 |
| Fe – O | $Fe_3O_4$ [14, 15,18] [4] or $PbFe_xV_{6-x}O_{11}$ [27] [5] | 120 65 | $LnO_{1-x}F_xFeAs$ [1,8,9] | ~50 |

[3] Unstable phases

[4] Verwey semiconductor – to – semiconductor transition

[5] p-type conductor – spin glass transition

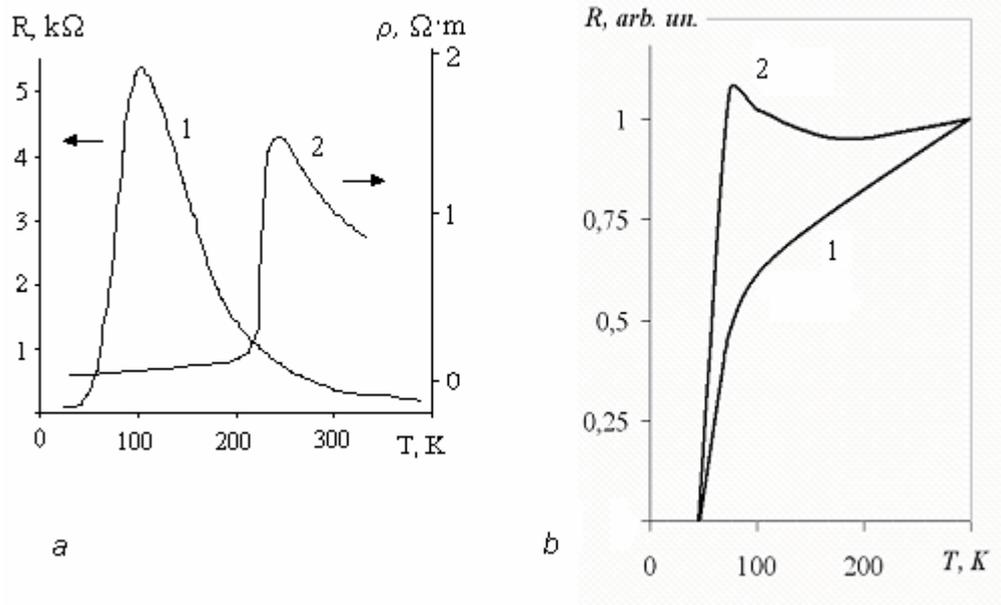

**Figure 1.** (a) Temperature dependences of resistance of the vanadium oxide sample subject to plasma treatment (1) from [7] and $La_{0.7}Ca_{0.3}MnO_3$ (2) (adopted from [13]); (b) Temperature dependences of the normalized resistance ($R(T)/R(300\ K)$) of the SC cuprates $TlCa_{1-x}Nd_xSr_2CuO_4$ at x=0.25(1) and x=0.75 (2) (after Edwards et al. [14])

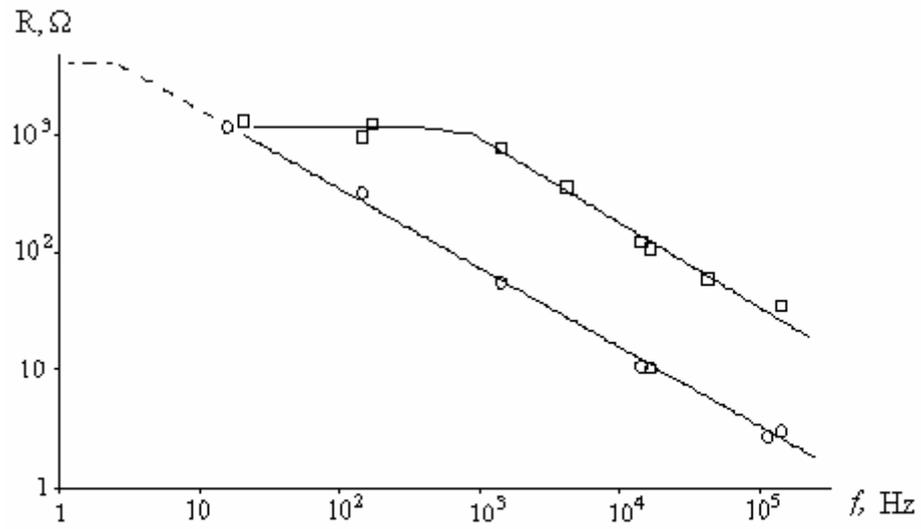

**Figure 2.** A.C. resistance of the vanadium oxide sample, subject to plasma treatment, as a function of frequency at $T = 20$ K (circles) and 293 K (squares).